\documentclass[a4paper]{article}
\pdfoutput=1 
\usepackage{subcaption}
\usepackage{INTERSPEECH2020}
\usepackage{makecell}
\usepackage[skip=3pt]{caption} 
\usepackage{multirow}

\title{Automatic Detection of Phonological Errors in Child Speech Using Siamese Recurrent Autoencoder}
\name{Si-Ioi Ng, Tan Lee}
\address{
  Department of Electronic Engineering, The Chinese University of Hong Kong}
 \email{siioing@link.cuhk.edu.hk, tanlee@ee.cuhk.edu.hk}

\begin{document}

\maketitle

\begin{abstract}
Speech sound disorder (SSD) refers to the developmental disorder
in which children encounter persistent difficulties in correctly pronouncing words. Assessment of SSD has been relying largely on trained speech and language pathologists (SLPs).
With the increasing demand for and long-lasting shortage of SLPs, automated assessment of speech disorder becomes a highly desirable approach to assisting clinical work. 
This paper describes a study on automatic detection of phonological errors in Cantonese speech of kindergarten children, based on a newly collected large speech corpus. The proposed approach to speech error detection involves the use of a Siamese recurrent autoencoder, which is trained to
learn the similarity and discrepancy between phone segments in the embedding space. Training of the model requires only speech data from typically developing (TD) children. To distinguish disordered speech from typical one, cosine distance between the embeddings of the test segment and the reference segment is computed. Different model architectures and training strategies are experimented. Results on detecting the $6$ most common consonant errors demonstrate satisfactory performance of the proposed model, with the average precision value from $0.82$ to $0.93$.

\end{abstract}

\section{Introduction}\label{intro} 
Children who suffer from speech sound disorder (SSD) commit persistent errors in producing certain speech sounds after the expected age of acquisition. Untreated children with SSD may experience social and academic difficulties, which impact their personal growth in the long term. Currently clinical assessment of SSD is carried out by qualified speech and language pathologists (SLPs) based on perceptual evaluation. The assessment can take various forms, including articulation test, conversation, story telling, etc. The result of each form of test reveals the severity and details of specific speech sound developmental problems. The assessment criteria are established and validated by experts. Timely diagnosis of SSD is crucial to effective treatment and rehabilitation. This is, however, hindered by the significant manpower shortage of SLPs globally. Methods of automatically detecting speech sound errors are highly desired to reduce the pressure on SLPs and benefit a large population of patients.

Child SSD detection is the task of distinguishing abnormal speech sound production from typical ones based on acoustic speech signals.
Possible approaches include template matching, statistical modeling and automatic speech recognition (ASR).
Given a limited amount of speech data, Yeung et al. \cite{yeung2017predicting} investigated an exemplar-based approach to evaluating English rhotic sounds in child speech. With a good amount of data for statistical modeling, Dudy et al. \cite{dudy2015pronunciation}\cite{dudy2018automatic} improved the goodness of pronunciation (GOP) \cite{witt2000phone} measure for pronunciation analysis in disordered child speech. Phonetic knowledge about common realizations of target phonemes was applied in the analysis. Similar approaches of knowledge incorporation were found in other works. 
In \cite{shahin2014comparison}, assessment of childhood apraxia of speech (CAS) was performed using constrained lattice in an ASR system. The lattice has the advantage that the type of mispronunciation could be beyond a binary decision. 
For each target word, the lattice was created according to expected mispronunciation rules. This approach was further extended in \cite{ward2016automated}, where phonological error patterns were identified via fine tuning of state transition weights between the correct and mispronounced phone sequences of a target word. 
However, the nature of being unpredictable in mispronunciation would challenge such systems, which rely on prior knowledge about the concerned errors.


In recent years fixed-dimension representation of speech has been applied widely to speech modeling and classification problems. Such representation encodes the information of variable-length speech segments in low-dimension vectors, which allow different segments to be compared and analyzed in the same embedding space. The similarity between segments can be evaluated by Euclidean distance, cosine distance, or other distance measures.
Many approaches have been proposed for extracting embedding from speech. In the present study, the use of sequence-to-sequence auto-encoder (AE) is investigated. It is a neural network model that encompasses an encoder-decoder architecture. The encoder converts the input sequence into a low-dimension embedding while the decoder aims to reconstruct from the embedding an output sequence that is the same as or closely related to the input. The applications of sequence-to-sequence AE are found in unsupervised spoken term discovery, query-by-example spoken term detection and speaker verification, etc. \cite{kamper2019truly}\cite{chung2016audio}\cite{lee2017discriminative}.

A common type of child SSD can be described as the desired phone, typically a consonant, being substituted by another phone. In this study, detection of such phonological errors is formulated as the problem of pairwise contrast between relevant phone segments, based on the embedding representations generated by an AE model. In terms of the network architecture, the AE is combined with a Siamese network, which is jointly trained to contrast the phone segments in the embedding space. Different model setups are evaluated first on test data of ``artificial'' substitution errors. Subsequently the proposed approach is applied to detect real phonological errors produced by children with SSD.

 \section{Background \& Speech Database}\label{corpus}





\subsection{Speech acquisition by Cantonese-speaking children} 
The present study is focused on Cantonese, a major Chinese dialect that is widely spoken in Hong Kong, Macau, Guangdong and Guangxi Provinces of Mainland China, as well as overseas Chinese communities. Cantonese is a monosyllabic and tonal language. Each Chinese character is pronounced as a single syllable carrying a lexical tone. A Cantonese syllable can be divided into an Initial part and a Final part. The Initial is a consonant while the Final could be a diphthong or comprise a vowel nucleus followed by a consonant coda (final consonant). There are a total of $19$ consonants, $11$ vowels and $11$ diphthongs in Cantonese. The present-day Cantonese uses over $700$ legitimate syllables (Initial-Final combinations). If the tone difference is taken into account, the number of distinct syllables exceeds $1,600$ \cite{bauer2011modern}\cite{lee2002spoken}. 
In this study, we focus on Cantonese spoken in Hong Kong. The target group of speakers is pre-school children in Hong Kong.

In \cite{so1995acquisition}, So and Dodd examined speech sounds of typically developing (TD) and Cantonese-speaking pre-school children. 
It was shown that children were able to acquire tones, most of the vowels and diphthongs by the age of $2$;$0$ (years;months). The acquisition of final consonants and initial consonants was achieved by the age of $4$;$6$ and $5$;$0$ respectively. To at el. \cite{to2013population} investigated acquisition of Hong Kong Cantonese by children aged $2$;$4$ to $12$;$4$. The study revealed a longer time required for speech sound acquisition. Vowels and diphthongs were acquired by $5$;$0$ and $4$;$0$ respectively, and all initial consonants were acquired by $6$;$0$.
In the process of speech sound acquisition, children may try to simplify a target speech sound by substituting it with other sounds. %
This is mainly due to the undeveloped motor skills for speech sound production. TD children gradually stop using the substitution sounds and return to typical pronunciation when they grow up. Nonetheless, some children would persist the substitution errors beyond the expected age of acquisition. The symptoms are referred to as phonological disorder and disordered children are recommended to seek treatment offered by SLPs. 

\begin{table}[th!]
\vspace{-2mm}
\centering
\caption{Statistics of speakers in available speech data}
\resizebox{\linewidth}{!}{%
\begin{tabular}{c|cccc}
\hline
\hline
Age (years;months)   & 3;0-3;11  & 4;0-4;11 & 5;0-5;11 & 6;0-6;11\\
\hline
Male, healthy     & $7$ & $26$ & $31$ & $14$ \\
Female, healthy    & $13$ & $33$ & $35$ & $20$ \\
Male, atypical    & $9$ & $9$  & $5$  & $1$ \\
Female, atypical    & $6$ & $9$ & $6$  & $0$ \\
\hline
\hline
\end{tabular}%
}
\label{tab:speaker_distribution}
\end{table}
\vspace{-3mm}

\subsection{Child Speech Database: CUCHILD}
A Cantonese child speech corpus named 
CUCHILD is used in the present study \cite{ng2020cuchild}. The corpus contains speech data collected from 
$1,986$ kindergarten children (aged $3$;$3$-$6$;$11$) in Hong Kong. All speakers use Cantonese as their first language (L1).
CUCHILD is designed to support acoustic modeling of Cantonese child speech and research on automatic assessment of SSD \cite{wang2018study}\cite{ng2018automated}. 
The speech material consists of a total of 
$130$ Cantonese words of $1$ to $4$ syllables in length, covering the $19$ consonants and $11$ most commonly used vowels. 
Speech recording was carried out in classrooms provided by the kindergartens. A digital recorder was located at $20$-$50$ centimeters in front of the children's mouth. Yet environmental noise such as reverberation, school bells, people walking around, etc. was unavoidable. To minimize effects of background noise, the gain and the position of recorders were adjusted manually.
Child speech was elicited via a picture naming task. Each word was also accompanied by a pictorial illustration. A research assistant showed the pictures one by one and guided the child to speak the intended words. 

All participants were assessed with the Hong Kong Cantonese Articulation Test (HKCAT) \cite{cheung2006hong}. The HKCAT is a standardized test for children which reflects the severity of developmental delay and the types of speech sound errors. Among all participants, $230$ children were found to have SSD. 

The speech data were collected recently and detailed work of data processing and annotation are still ongoing, The present study makes use of a subset of the whole corpus, which covers the recordings from $233$ child speakers. The data was manually annotated and segmented into child speech and research assistants' speech. Spoken words manifesting SSD were labelled manually by SLPs. The syllable-level orthographic transcriptions were manually verified. Table~\ref{tab:speaker_distribution} summarizes the speaker information in our dataset.

\begin{figure}[t]
  \setlength\belowcaptionskip{-0.8\baselineskip}
  \centering
  \includegraphics[width=0.85\linewidth]{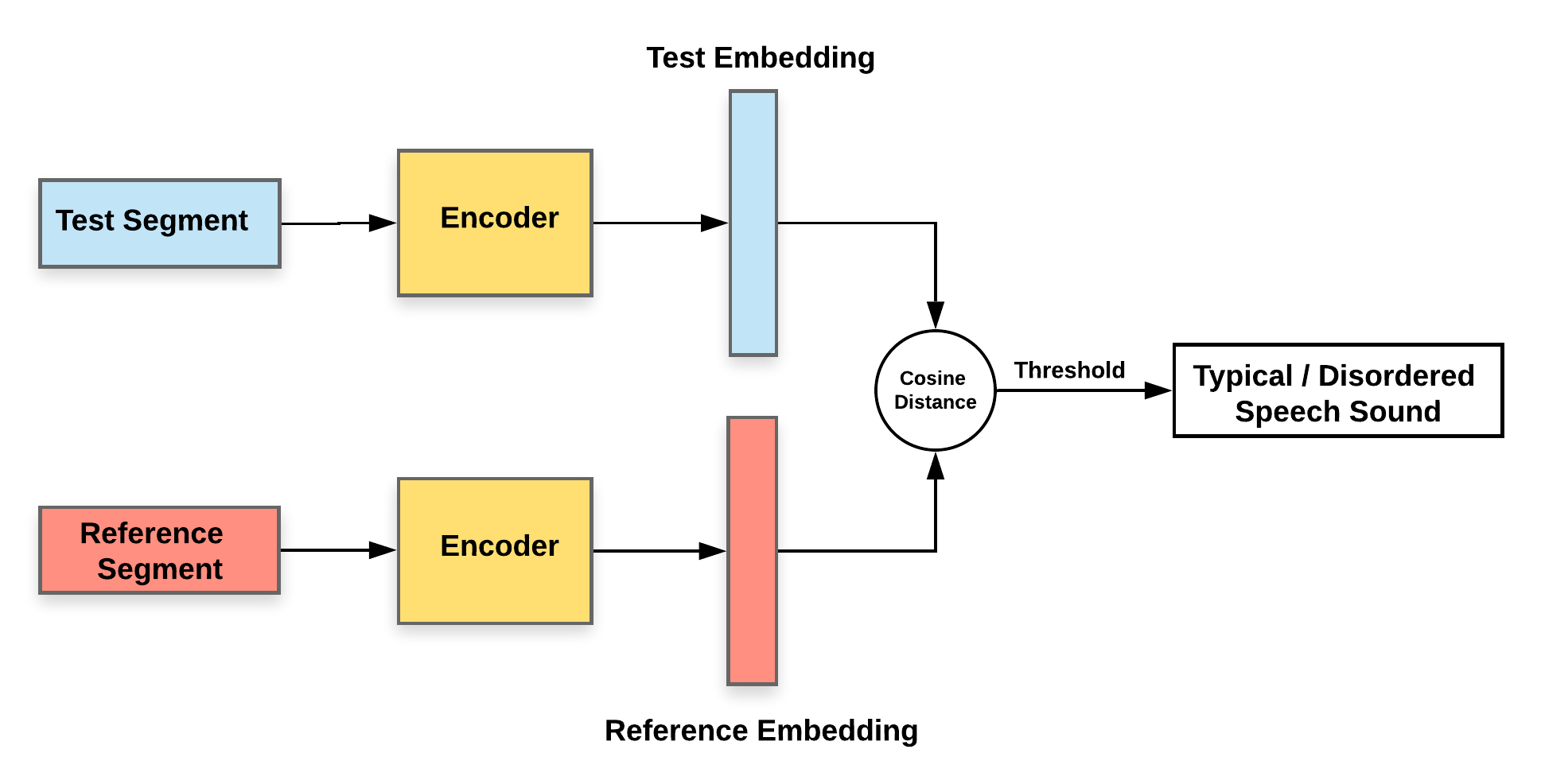}
  \caption{Speech sound disorder (SSD) detection system.} 
  \label{fig:system_design}
  \vspace{-3mm}
\end{figure}

\section{The Proposed System of SSD Detection}\label{system}

\subsection{SSD detection system} 

In clinical assessment of SSD, the child is guided to speak a list of test words. The responsible speech pathologist observes the speech production and decides if the child makes errors on specific parts of the words. The judgement depends highly on the clinician's experience in differentiating atypical speech sounds from typical ones.



Towards automated assessment of SSD, the proposed system aims to determine whether a phonological error occurs in a test speech segment. The test segment contains a specific phoneme as part of a test word spoken by the child. 
To detect the error, we may choose one or multiple reference segments of the expected speech sound to compare with the test segment in a pairwise manner. Using multiple reference segments is preferred as they can represent the deviation of the expected speech sound.
As illustrated in Figure \ref{fig:system_design}, the comparison is in an embedding space and all embeddings are extracted by the encoder obtained from the trained Siamese RAE model. The cosine distance is computed for each pair of embedding. The binary decision is based on a pre-defined threshold. If the score is above the threshold, the test segment is classified as typical pronunciation. Otherwise it is a disordered pronunciation.

The present study is focused on a set of initial consonants in Cantonese, which are considered as reliable markers for child speech acquisition.  Details of the model are described in the following sections.



\subsection{Recurrent autoencoder}
A recurrent autoencoder (RAE) model is used to generate a compact representation of phone segment. This representation is referred to as the embedding. The RAE converts variable-length phone segments into fixed-dimensional embedding vectors, on which distance or similarity measure could be applied straightforwardly.
The RAE has three components. The encoder receives an input sequence. The hidden state of the encoder's last layer reaches the linear layer and generates the embedding, which is passed to the decoder to construct the output sequence. The RAE is trained such that the embedding is adequate for reconstructing a certain type of target output. One common choice of the target output sequence is to make it equal to the input sequence. This can be achieved by minimizing the mean squared error (MSE) loss in the training of encoder and decoder networks. For the input sequence $S = [\: \boldsymbol{x_{1}, x_{2}, x_{3},...,x_{T}}\:]$, the MSE loss is given as,
\begin{equation} \label{MSE}
L_{mse}=\sum_{t=1}^{T} \| \boldsymbol{x_{t}} - D_{t}(E(S)) \| ^2    
\end{equation}
where $D_{t}(\cdot)$ refers to the decoder output at time step $t$ and $E(\cdot)$ denotes the last hidden layer output of the encoder, while $T$ is the length of input sequence.

The RAE model is also commonly applied with a weakened input-output relation, i.e., without requiring the decoder to perform exact recovery of the input sequence. Such design aims at sharing mutual information between non-identical but closely related training segments \cite{kamper2019truly}. 
This type of RAE is known as the correspondence RAE (Cor-RAE). In this work, the Cor-RAE model is trained using speech segments carrying the same phoneme.
Consider a pair of segments $S_1 = [\: \boldsymbol{x^{(1)}_{1}, x^{(1)}_{2}, x^{(1)}_{3},...,x^{(1)}_{T_1}}\:]$ and $S_2 = [\: \boldsymbol{x^{(2)}_{1}, x^{(2)}_{2}, x^{(2)}_{3},...,x^{(2)}_{T_2}}\:]$ from the same phoneme category, the MSE loss for the training of Cor-RAE is,
\begin{equation} \label{MSE_Cor}
L_{mse}=\sum_{t=1}^{T_2} \| \boldsymbol{x^{(2)}_{t}} - D_{t}(E(S_1)) \| ^2    
\end{equation}
where $T_1$ and $T_2$ denote the lengths of $S_1$ and $S_2$ respectively. 



\begin{figure}[t!]
  \setlength\belowcaptionskip{-0.8\baselineskip}
  \centering
  \includegraphics[width=0.85\linewidth]{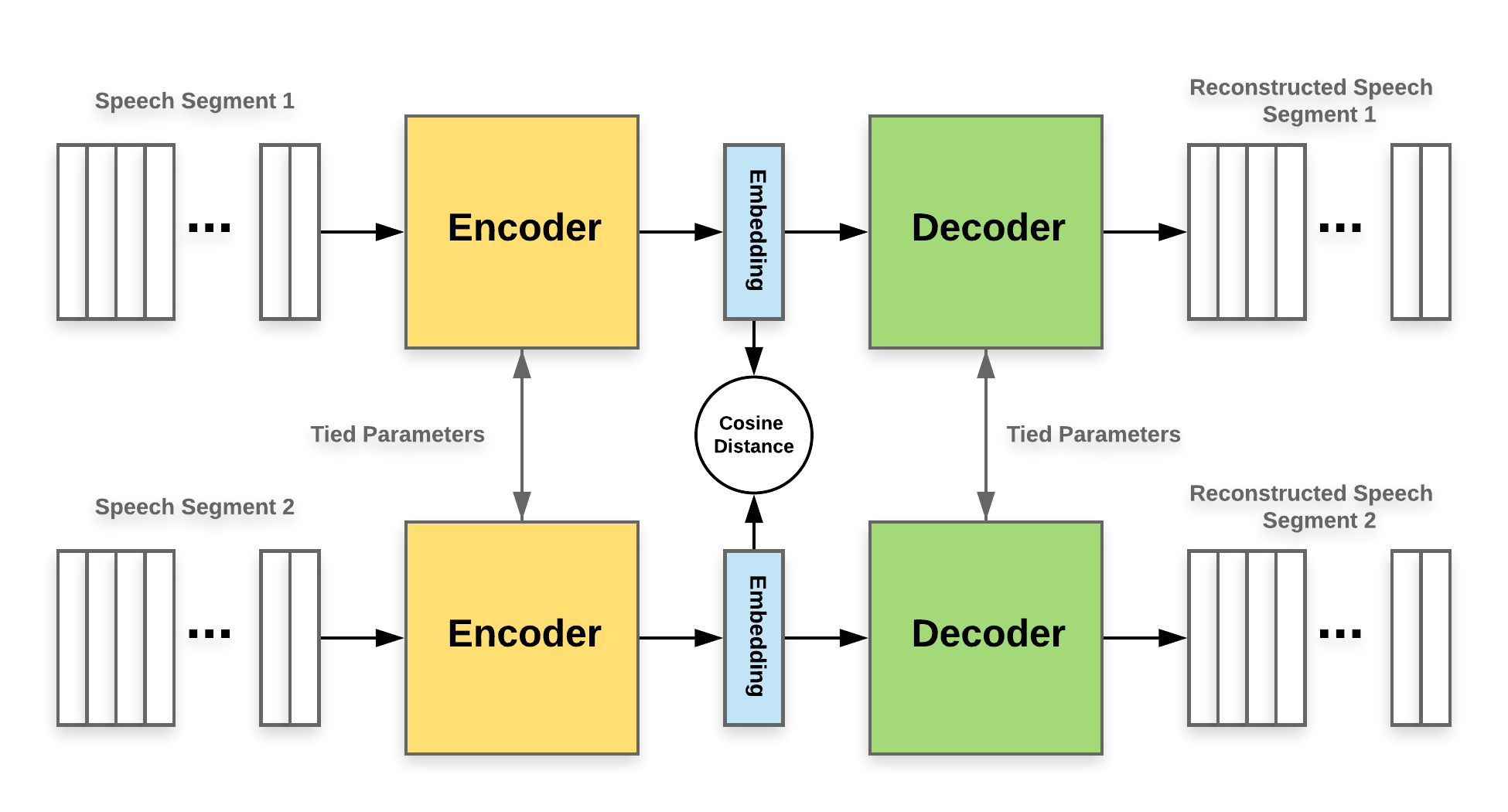}
  \caption{Siamese network architecture.}
  \label{fig:RAE}
  \vspace{-2mm}
\end{figure}

\subsection{Siamese recurrent autoencoder}
As discussed earlier, the task of phonological error detection is formulated as a process of contrasting a test segment against the target phonemes. This process is realized with a Siamese network. It consists of two identical neural networks with shared parameters, which process two input representations in parallel. By inserting the Siamese loss in training, the network parameters are optimized to learn the similarity between the input representations. Our implementation of the Siamese RAE follows the work in \cite{zhu2018siamese}, where the loss is computed with a pair of embeddings extracted from the RAE, as shown in Figure \ref{fig:RAE}. 

Two types of Siamese loss are considered and compared in this work. The first one is the contrastive loss, which is expressed as,



\begin{equation} \label{contrast_loss}
L_{c}=\dfrac{1}{2}y*d + \dfrac{1}{2}(1-y)*{max(0,m-d)},
\end{equation}
where $d=1-cos(\boldsymbol{z_{1},z_{2}})$, and $\boldsymbol{z_{1},z_{2}}$ are the pair of embeddings representing two input speech segments. Both embeddings are generated by the encoder in the Siamese RAE.

The other type of loss function is the triplet loss defined as,
\begin{equation} \label{triplet_loss}
L_{t}={max(0,m+d_{ap}-d_{an})},
\end{equation}
The loss function involves three embeddings as input, which include an anchor $z_a$, a positive sample $z_p$ and a negative sample $z_n$. $d_{ap}$ and $d_{an}$ are the cosine distances of the anchor-positive pair and the anchor-negative pair respectively, and $d_{ap}=0.5*(1-cos(\boldsymbol{z_a,z_p}))$.

The overall objective function for Siamese RAE training combines the MSE loss and the contrastive/triplet loss as,
\begin{equation} \label{mse_contrast_loss}
\begin{split}
L_{mse,c}& = (1-w)*L_{c} +  w*\dfrac{L_{mse1}+Loss_{mse2}}{2}
\end{split}
\end{equation}
\begin{equation} \label{mse_triplet_loss}
\begin{split}
L_{mse,t} & = (1-w)*L_{t} +  w*\dfrac{L_{mse\_a}+L_{mse\_p}+L_{mse\_n}}{3}
\end{split}
\end{equation}
where $w$ is a scalar weight to balance the reconstruction loss and similarity loss. 


\section{Experiments and Results}\label{exp}

\subsection{Data pre-processing}\label{ali}
Consonant segments in TD and atypical child speech were extracted automatically by forced alignment with GMM-HMM triphone models. The triphone models were trained with speech data from $80$ TD children of age $5;0$ - $6;11$ among the $233$ speakers as summarized in Table 1. TD children in this age range are expected to make few mistakes in speech production and their speech are considered to be free of SSD problems.
Acoustic features for GMM-HMM training consist of $13$-dimensional Mel-frequency cepstral coefficients (MFCC) and their first- and second-order derivatives extracted every $0.01$ second.
For triphone model training, linear discriminant analysis (LDA), semi-tied covariance (STC) transform and feature space Maximum Likelihood Linear Regression (fMLLR) were applied \cite{duda2012pattern}\cite{gales1999semi}\cite{gales1998maximum}. 
With a basic syllable pronunciation dictionary, an error rate of $17.35$\% was achieved on the task of free-loop syllable recognition with test speech from $15$ unseen TD children in the same age range. 
Forced alignment was applied to the speech data shown in Table \ref{tab:speaker_distribution} according to the canonical pronunciations of the $130$ test words. Feature extraction, acoustic model training and forced alignment were all carried out with the Kaldi speech recognition toolkit \cite{povey2011kaldi}. As a result, a pool of consonant segments were extracted and they  
were divided into different subsets as shown in Table \ref{tab:data_RAE}.

\begin{table}[h!]
\caption{Summary of data (phone segments) for training and evaluation of the RAE model.}
\centering
\resizebox{\linewidth}{!}{%
\begin{tabular}{ |c||c|c|c| } 
 \hline
\textbf{Name of subset} & \textbf{Clinical group} & \textbf{Age range} & \textbf{No. of segments} \\
\hline
Training & TD & 5;0 - 6;11 & $17400$ \\ 
\hline
Reference & TD &  5;0 - 6;11 & $4000$ \\
\hline
Development & TD & 5;0 - 6;11 & $3500$ \\
\hline
\hline
Test1 & TD & 3;0 - 4;11 & $21000$ \\
\hline
Test2 & TD \& Disordered & 3;0 - 6;11 & $706$ \& $726$ \\
\hline
\end{tabular}
}
\label{tab:data_RAE}
\vspace{-6mm}
\end{table}

\subsection{Training of the Siamese RAE}\label{tr_rae} 
In this study, the gated recurrent units (GRU) are adopted as the recurrent neural network architecture in the Siamese RAE  \cite{audhkhasi2017end}. 
The input representations are $40$ dimensional Filter-bank features, with mean and variance being globally normalized.
The training phone segments are paired randomly. 
A training target '$1$' is assigned to the pairs of same-class segments 
, and '$0$' assigned to pairs of segments from different classes. 
The encoder and decoder networks both consist of $3$ hidden layers and $400$ hidden units. The embeddings are L2-normalized. 
Both the Siamese RAE and Siamese Cor-RAE models are trained by the Adam optimizer \cite{kingma2014adam} with a batch size of $256$, a learning rate of $10^{-4}$, weight decay of $10^{-5}$ and for $50$ epochs. Training of the Siamese Cor-RAE starts with a pre-trained standard Siamese RAE model. 
The margin is $0.9$ for the contrastive loss and $0.25$ for the triplet loss. A loss weight of $0.5$ is applied to both loss functions. 
The training processes are implemented with PyTorch \cite{paszke2019pytorch}.

\begin{figure}[t!]
  \setlength\belowcaptionskip{-0.8\baselineskip}
  \centering
  \includegraphics[width=0.68\linewidth]{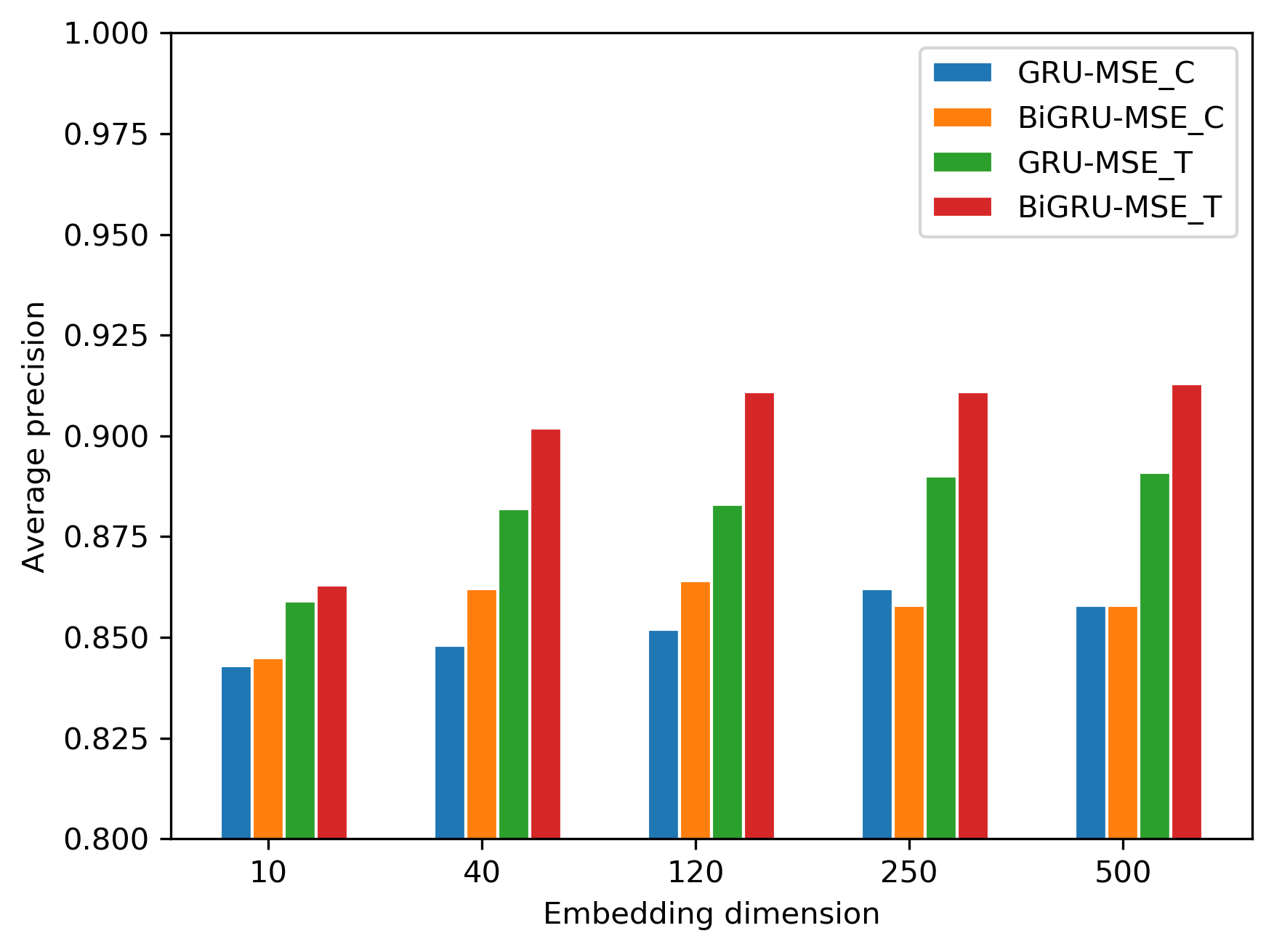}
  \caption{Performance on the development set.}
  \label{fig:AP_Dim}
  \vspace{-3mm}
\end{figure}

Different embedding sizes, loss functions (contrastive vs. triplet) and Siamese RAE model designs are evaluated on the development set. The evaluation is carried out on the same-different discriminability task as described in \cite{carlin2011rapid}. Given a pair of test segments $(p_{i},p_{j}), i \neq j$, $p_{i}$ and $p_{j}$ are declared to contain the same phoneme if the embedding distance $d \leq \tau$, where $\tau$ is the decision threshold.
Each development segment is randomly paired with a segment from the reference dataset. 
The segment pairs assigned with '$0$' are regarded as artificial substitution errors, in which the target phone is substituted by another phone. The cosine distance is computed for each segment pair. The average precision (AP) is used as the evaluation metric of system performance. The value of AP is obtained from the precision-recall (PR) curve, which portrays the system performance across varying decision thresholds.
The results in terms of AP are shown as in Figure \ref{fig:AP_Dim}. Overall, using the triplet loss and bi-directional network structure (BiGRU-MSE\_T) leads to better performance on the same-different task. 
In the following experiments, this setting with an embedding size of $120$ is used.

\subsection{Performance evaluation on artificial errors} \label{phone_dis}
\begin{figure}[t!]
  \setlength\belowcaptionskip{-0.8\baselineskip}
  \centering
  \includegraphics[width=0.68\linewidth]{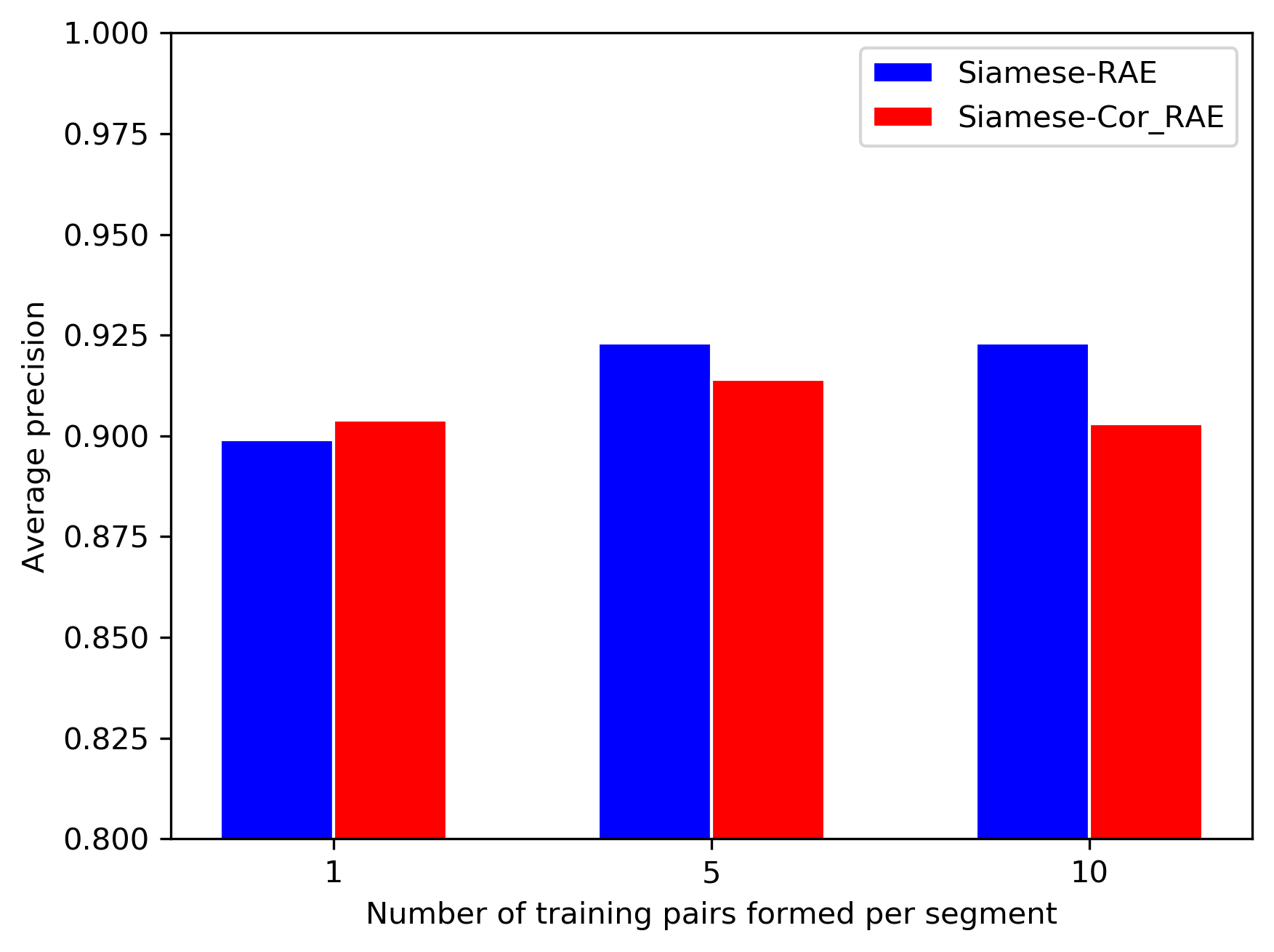}
  \caption{Performance on artificial errors.}
  \label{fig:Arti}
  \vspace{-3mm}
\end{figure}

\vspace{-1mm}
In this part, the \textit{Test1} dataset in Table \ref{tab:data_RAE} is used to evaluate the performance of the Siamese RAE and Siamese Cor-RAE. Each test segment in \textit{Test1} is paired up with a segment randomly selected from the reference set. The results in terms of AP are reported as in Figure \ref{fig:Arti}. In the figure we also compare different training strategies in which each segment in the training dataset is used to form $1$, $5$ and $10$ training pairs with other training segments.
It can be seen that the conventional Siamese RAE consistently outperforms the Siamese Cor-RAE. 
The change of training pairs shows a noticeable impact on the performance level. The results imply that it is beneficial to use more training pairs. However, a large number of training pairs would not yield further improvement, in particular for the Siamese Cor-RAE. 
It seems the learning of mutual information shared across short segments from the same phoneme in Siamese Cor-RAE does not 
work as successfully as where sub-word and word level speech units are used \cite{kamper2019truly}\cite{renshaw2015comparison}. This could be caused by the short duration of speech units or hyperparameter settings of the model. More works are required to draw definitive conclusions. 


\subsection{Performance evaluation on real errors}
The Siamese RAE with Bi-GRU trained with $5$ training pairs, i.e., the best performing model shown in Figure \ref{fig:Arti}, is evaluated on the task of detecting real phonological errors with the \textit{Test2} dataset. 
The test data cover the $6$ most common error patterns, which concern the Cantonese consonants /f/, /k/, /s/, /k\textsuperscript{h}/, /t\textsuperscript{h}/ and /p\textsuperscript{h}/. The errors are made on the phonological processes of stopping (e.g. /f/ to /p/), fronting (e.g. /k/ to /t/), deaspirtation (e.g. /k\textsuperscript{h}/ to /k/) and affrication (e.g. /s/ to /ts/) etc. They are caused mainly by incorrect place or manner of articulation. It should be noted that children with SSD often had incomplete speech sound inventories and the errors were not limited to these common patterns.


\begin{table}[h!]
\vspace{-3mm}
\centering
\caption{AP on real consonant errors.}
\resizebox{\linewidth}{!}{%
\begin{tabular}{ |c|c|c|c| } 
 \hline
 \textbf{Consonant} & \textbf{Error Pattern} & \textbf{\makecell{No. of  Consonant Segments \\ \{Disordered, Typical, Reference\}}} & \textbf{AP} \\
 \hline
 /f/ & \makecell{Stopping} & \{$65$,$93$,$153$\} & $0.898$ \\ 
 \hline
 /k/ & \makecell{Fronting} & \{$52$,$137$,$409$\} & $0.824$ \\
 \hline
 /s/ & \makecell{Affrication, Stopping} & \{$178$,$183$,$294$\} & $0.917$\\
 \hline
 /k\textsuperscript{h}/ & \makecell{Deaspiration, Fronting} & \{$141$,$102$,$169$\} & $0.938$\\
 \hline
 /t\textsuperscript{h}/ &\makecell{Deaspiration, Backing} & \{$205$,$115$,$220$\} & $0.861$\\
 \hline
 /p\textsuperscript{h}/ & \makecell{Deaspiration} & \{$85$,$76$,$126$\} & $0.921$\\
 \hline
\end{tabular}
}
\label{tab:disorder_eva}
\vspace{-3mm}
\end{table}

Each test segment is compared with all reference segments carrying the same consonant. The average cosine distance is computed. The results in terms of AP are shown as in Table \ref{tab:disorder_eva}. 
The highest AP is achieved on the detection of atypical aspirated consonant /k\textsuperscript{h}/ sound, while the performance of detecting unaspirated consonant /k/ is the worst among all test patterns. This suggests that unaspirated consonants may not be reliably detected in automatic assessment of SSD. It was noted that the Siamese Cor-RAE did not yield good performance. The use of the correspondence model for SSD detection with higher-level speech units (e.g. syllable or word level) will be investigated in our future work.

%


\vspace{-1mm}
\section{Conclusion}\label{conclusions}
An approach to automatic detection of phonological errors in child speech has been investigated and evaluated with both artificial and real speech sound errors. It has been shown that the proposed Siamese Recurrent Auto-encoder model is able to learn compact representations from variable-length speech segments, which are effective in distinguishing erroneous segments from correct ones. Specifically, for the $5$ most common consonant errors in Cantonese, the achieved values of average precision range from $0.82$ to $0.93$. These results reveal the good potential of applying the proposed approach to automatic assessment of speech sound disorder in real-world settings. Future work will include the incorporation of clinical knowledge in the model design and the discovery of domain knowledge through acoustical analysis of child speech.




\bibliographystyle{IEEEtran}

\bibliography{me}

\end{document}